\documentclass[aps, prd, superscriptaddress, nofootinbib, preprintnumbers]{revtex4-2}
\usepackage[colorlinks=true]{hyperref}
\usepackage{xcolor}
\hypersetup{colorlinks=true, citecolor=blue, urlcolor=blue, linkcolor=blue}
\usepackage{amsmath,amssymb}
\usepackage[final]{graphicx}
\usepackage{subcaption}
\usepackage[belowskip=0.5pt,aboveskip=0.5pt]{caption}
\usepackage{float}
\usepackage{adjustbox}
 \usepackage {ulem}

\begin{document}

\title{$D_{(s)}-$ mesons semileptonic form factors in the 4-flavor holographic QCD}

\author{Hiwa A. Ahmed}
\email{hiwa.a.ahmed@mails.ucas.ac.cn}
\affiliation{School of Nuclear Science and Technology, University of Chinese Academy of Sciences, Beijing, 100049, P.R. China}
\affiliation{Department of Physics, College of Science, Charmo University, 46023, Chamchamal, Sulaymaniyah, Iraq}

\author{Yidian Chen}
\email{chenyidian@hznu.edu.cn}
\affiliation{School of Physics, Hangzhou Normal University, Hangzhou, 311121, P.R. China}

\author{Mei Huang}
\email{huangmei@ucas.edu.cn}
\affiliation{School of Nuclear Science and Technology, University of Chinese Academy of Sciences, Beijing, 100049, P.R. China}

\date{\today}

\begin{abstract}

We investigate semileptonic form factors of $D_{(s)}$ meson from a modified soft-wall 4-flavor holographic model.  The model successfully reproduces the masses and decay constants of various mesons, including $\rho$, $K^*$, $D^*$, $D_s^*$, $a_1$, $K_1$, $f_1$, $D_1$,$D_{s1}$, $\pi$, $K$, $\eta$, $D$, and $D_s$. Moreover, we study the semileptonic decay processes $D^{+} \to (\pi, K, \eta) l^{+} \nu_{l}$ and $D_{s}^{+} \to ( K, \eta) l^{+} \nu_{l}$, associated with the vector meson exchange, as well as $D_{(s)}^{+} \to K^{} l^{+} \nu_{l}$, associated with the vector and axial vector meson exchange. The form factors $f_{+}(q^{2})$ for $D \to\pi$ and $D_{(s)}\to K$ decays agree excellently with experimental and lattice data, outperforming other theoretical approaches. The $f_{+}(q^{2})$ form factor for $D^{+} \to \eta $ is compatible with experimental data, while a slight discrepancy is observed for $D_{s}^{+} \to \eta $ at large $q^{2}$. Additionally, we predict the vector form factors $V(q^{2})$ and $A_{1}(q^{2})$ for $D \to K^{}$ and $D_{s} \to K^{}$ decays, respectively. The results agree well with other approaches and lattice data at maximum recoil ($q^{2}=0$).

\end{abstract}


\maketitle

\section{Introduction}

Semileptonic weak decays of mesons play a vital role in our comprehension of the standard model (SM) as they provide the most direct way to determine the Cabibbo-Kobayashi-Maskawa (CKM) matrix \cite{Cabibbo:1963yz,Kobayashi:1973fv} elements from experimental data. In particular, semileptonic $D_{(s)}$ meson decays offer a valuable avenue for investigating the interactions within the charm sector, where by measuring the decay rates, it becomes possible to directly determine the CKM matrix elements $|V_{cd}|$ and $|V_{cs}|$. For instance, the values of $|V_{cd}|$ and $|V_{cs}|$ are found from the measurements of the decays $D \to \pi l \mu_{l}$ and $D \to K l \mu_{l}$, respectively, by Belle \cite{Belle:2006idb}, BaBar \cite{BaBar:2007zgf,BaBar:2014xzf}, CLEO \cite{CLEO:2009svp}, and BESIII \cite{BESIII:2015tql} collaborations. It is worth noting that extracting the CKM matrix elements is not straightforward, rather, it includes the nonperturbative strong effects appearing in the transition from the initial state to the final state, which is parameterized by the hadronic invariant form factors. More recently, The BESIII collaboration reports several semileptonic weak decays, such as  $D^{+} \to K^{-} \pi^+ e^+ \nu_e$ \cite{BESIII:2015hty}, $D^+_{s}\rightarrow K^0 e^+\nu_e$ and $D^+_{s}\rightarrow K^{*0} e^+\nu_e$ Decays in Ref. \cite{BESIII:2018xre}, $D_s^+ \rightarrow \eta^{(\prime)} e^+ \nu_e$ in Ref. \cite{BESIII:2019qci}, and $D^+ \rightarrow \eta\mu^+\nu_\mu$ in \cite{Ablikim:2020hsc}. Since the semileptonic decays include the nonperturbative hadronic form factors, one can not use the direct quantum chromodynamics (QCD), and one needs a nonperturbative method to carry out the calculations, see Ref. \cite{Soni:2018adu} for listing the theoretical approaches.

Apart from the other nonperturbative approaches, a holographic QCD model was applied to describe the structure of the hadrons. Based on the anti-de Sitter/conformal field theory (AdS/CFT) correspondence discussed in Refs. \cite{Maldacena:1997re,Witten:1998qj}, a bottom-up holographic QCD model at low energy was established in the works of Refs. \cite{Erlich:2005qh,Karch:2006pv,Gubser:2008ny,Gubser:2008yx,Grefa:2021qvt,Gursoy:2007cb,Gursoy:2007er,Chen:2022goa,Li:2012ay,Li:2013oda,MartinContreras:2021yfz,Contreras:2021epz,Cao:2022csq,Chen:2022pgo}. They started from QCD and constructed a five-dimensional dual with the features of the dynamical chiral symmetry breaking. Since in the two-flavor system, the masses of the up and down quarks are small, and an $SU(2)$ flavor symmetry is preserved. However, in the case of the extension of the model to three flavors \cite{Abidin:2009aj} and four flavors \cite{Ballon-Bayona:2017bwk, Momeni:2020bmy,Momeni:2022gqb, Chen:2021wzj,Ahmed:2023zkk}, the flavor symmetry is broken, especially in the case of including Charm quark. The first attempt to study the semileptonic decays had been done in Ref. \cite{Abidin:2009aj}, where the $K_{l3}$ form factors that describe the decays $K \to \pi l \nu_{l}$ was calculated. More recently, the semileptonic $D$ meson decays to the vector, axial vector, and scalar mesons investigated in the hard-wall holographic approach \cite{Momeni:2022gqb}.

In the present work, we use 4-flavor bottom-up holographic framework to study the semileptonic decays. In the original soft-wall holographic model \cite{Karch:2006pv}, the quark condensate is proportional to the quark mass, which is in contradiction with QCD, so to overcome the issue, a higher order potential is added to the 5D action \cite{Gherghetta:2009ac}. Therefore, we adopt the modified 4-flavor soft-wall model \cite{Ahmed:2023zkk} instead of the soft-wall model. Following, we proceed by calculating the  masses and decay constants of the $\pi$, $K$, $\eta$, $D$, $D_{s}$, $\rho$, $K^*$, $\omega$, $D^{*}$, $D_{s}^{*}$, $a_{1}$, $K_{1}$, $f_{1}$, $D_{1}$, and $D_{s1}$ mesons in the ground state. Furthermore, we compute the form factors of the semileptonic decays $D^{+} \to (\pi, K, \eta, K^{*}) l^{+} \nu_{l}$ and $D_{s}^{+} \to ( K, \eta, K^{*}) l^{+} \nu_{l}$ which induced by the decay of the charm quark to light quark, $c \to d(s) l \nu_{l}$. Due to the fact that the maximum-recoil form factors are essential to extract the CKM matrix elements, and they are also observable in the experiment, we compare our determined value with the experimental and lattice QCD data.

This work is organized as follows. In section \ref{modelrevisit}, we revisit the formalism of the modified soft-wall holographic QCD model for $N_{4}$ flavor and derive the equations of motion. In section \ref{semi-form}, we describe the three-point interactions and deduce the semileptonic form factors from the three-point functions obtained from the cubic-order 5D action. A detailed comparison of the numerical results  with the experimental data, lattice QCD, and other theoretical approaches are provided in section \ref{result}. Finally, we briefly conclude our work in section \ref{conclusion}.

\section{The 5D action and equations of motion}
\label{modelrevisit}

In this section, we revisit the formalism of the four flavors of soft-wall holographic QCD model \cite{Chen:2021wzj,Ahmed:2023zkk}. The five-dimensional metric defined in the AdS space is given by, 

\begin{equation}
ds^{2}= g_{MN} dx^{M} dx^{N} =\frac{1}{z^{2}} \left( \eta_{\mu \nu} dx^{\mu} dx^{\nu} + dz^{2}\right),
\end{equation}
where $\eta_{\mu \nu}=\operatorname{diag}\left[-1, 1,1, 1\right]$ is the four-dimensional metric in the Minkowski space, and $z$ is the fifth dimension and has an inverse energy scale. Note that the Latin indices $M$ and $N$ run from $0,1,2,3,4$, and the Greek indices are defined as $\mu$, $\nu =0,1,2,3$. According to the holographic model, there is a correspondence between the 4D operators and Corresponding 5D gauge fields \cite{Erlich:2005qh}. The operators and corresponding gauge fields incorporated in the chiral dynamics are defined by 

\begin{equation}
\begin{aligned}
&J_{R/L \mu}^a=\bar{\psi}_{q R/L} \gamma_\mu t^a \psi_{q R/L} \to R_{\mu}^{a}/L_{\mu}^{a}\\
& J^{S}=\bar{\psi}_{q L} \psi_{q R} \to X.
\end{aligned}
\end{equation}
where $J_{R/L \mu}^a$ is a right/left-handed currents which correspond to the $R_{\mu}^{a}$ and $L_{\mu}^{a}$ gauge fields, and the quark bilinear $\bar{\psi}_{q L} \psi_{q R}$ correspond to the complex scalar fields $X$. Note that, $t^{a}$ with $a=1,2, ..., N_{f}^{2}-1$ are the generators of the $SU(N_{f})$ group. Th general five-dimensional action is written as

\begin{equation}
\begin{aligned}
S_{M} &=-\int_{\epsilon}^{z_{m}} d^{5} x \sqrt{-g} e^{-\phi} \operatorname{Tr}\left\{\left(D^{M} X\right)^{\dagger}\left(D_{M} X\right) + M_{5}^{2}|X|^{2} -\kappa |X|^4\right.\\
&\left.+\frac{1}{2 g_{5}^{2}}\left(V^{M N} V_{M N}+A^{M N} A_{M N}\right)\right\},
\end{aligned}
\label{action}
\end{equation}
where $D^{M} X=\partial_{M} X - i \left[V_{M}, X\right] - i \{A_{M}, X\}$ is the covariant derivative of the scalar field $X$, $M_{5}^{2}=(\Delta - p)(\Delta + p -4) =-3$ by taking the conformal dimension of the scalar field operator $\Delta=3$ and $p=0$, $\kappa$ is a dimensionless parameter which can be determined, and $\epsilon$ and $z_{m}$ are the UV and IR limit of the model. The coupling constant $g_{5}$ is related to the number of color and defined $g_{5} = 2 \pi $ for $N_{c}=3$  (\cite{Erlich:2005qh}). The gauge field strength $V_{M N}$ and $A_{M N}$ are defined by 

\begin{equation}
\begin{aligned}
&V_{M N}=\partial_{M} V_{N}-\partial_{N} V_{M}-i\left[V_{M}, V_{N}\right] -i\left[A_{M}, A_{N}\right], \\
&A_{M N}=\partial_{M} A_{N}-\partial_{N} A_{M}-i\left[V_{M}, A_{N}\right] -i\left[A_{M}, V_{N}\right], \\
\end{aligned}
\end{equation}
where the vector and axial vector fields are written in terms of the right- and left-handed gauge fields as $V_{M}=\frac{1}{2}(L_{M} + R_{M})$ and $A_{M}=\frac{1}{2}(L_{M} - R_{M})$, respectively. The fields $V_{M}$, and $A_{m}$ can be expanded to $V_{M}^{a} t^{a}$, and $A_{m}^{a} t^{a}$, respectively, and the generators satisfy $Tr(t^{a} t^{b})=\frac{1}{2} \delta^{ab}$. The vector, axial and psudoscalar fields are described by $4 \times 4$ matrices,

\begin{equation}
\begin{aligned}
& V=V^a t^a=\frac{1}{\sqrt{2}}\left(\begin{array}{cccc}
\frac{\rho^0}{\sqrt{2}}+\frac{\omega^{\prime}}{\sqrt{6}}+\frac{\psi}{\sqrt{12}} & \rho^{+} & K^{*+} & \bar{D}^{* 0} \\
\rho^{-} & -\frac{\rho^0}{\sqrt{2}}+\frac{\omega^{\prime}}{\sqrt{6}}+\frac{\psi}{\sqrt{12}} & K^{* 0} & D^{*-} \\
K^{*-} & \bar{K}^{* 0} & -\sqrt{\frac{2}{3}} \omega^{\prime}+\frac{\psi}{\sqrt{12}} & D_s^{*-} \\
D^{* 0} & D^{*+} & D_s^{*+} & -\frac{3}{\sqrt{12}} \psi
\end{array}\right), 
\end{aligned}
\end{equation}
\begin{equation}
\begin{aligned}
& A=A^a t^a=\frac{1}{\sqrt{2}}\left(\begin{array}{cccc}
\frac{a_1^0}{\sqrt{2}}+\frac{f_1}{\sqrt{6}}+\frac{\chi_{c 1}}{\sqrt{12}} & a_1^{+} & K_1^{+} & \bar{D}_1^0 \\
a_1^{-} & -\frac{a_1^0}{\sqrt{2}}+\frac{f_1}{\sqrt{6}}+\frac{\chi_{c 1}}{\sqrt{12}} & K_1^0 & D_1^{-} \\
K_1^{-} & \bar{K}_1^0 & -\sqrt{\frac{2}{3}} f_1+\frac{\chi_{c 1}}{\sqrt{12}} & D_{s 1}^{-} \\
D_1^0 & D_1^{+} & D_{s 1}^{+} & -\frac{3}{\sqrt{12}} \chi_{c 1}
\end{array}\right),
\end{aligned}
\end{equation}
\begin{equation}
\begin{aligned}
& \pi=\pi^a t^a=\frac{1}{\sqrt{2}}\left(\begin{array}{cccc}
\frac{\pi^0}{\sqrt{2}}+\frac{\eta}{\sqrt{6}}+\frac{\eta_c}{\sqrt{12}} & \pi^{+} & K^{+} & \bar{D}^0 \\
\pi^{-} & -\frac{\pi^0}{\sqrt{2}}+\frac{\eta}{\sqrt{6}}+\frac{\eta_c}{\sqrt{12}} & K^0 & D^{-} \\
K^{-} & \bar{K}^0 & -\sqrt{\frac{2}{3}} \eta+\frac{\eta_c}{\sqrt{12}} & D_s^{-} \\
D^0 & D^{+} & D_s^{+} & -\frac{3}{\sqrt{12}} \eta_c
\end{array}\right) .
\end{aligned}
\end{equation}

Additionally, The complex scalar field in Eq. \eqref{action} is expressed by 

\begin{equation}
X=e^{i \pi^{a} t^{a}} X_{0} e^{i \pi^{a} t^{a}}
\end{equation}
where $X_{0}=\frac{1}{2}\operatorname{diag}\left[v_{l}(z), v_{l}(z), v_{s}(z), v_{c}(z)\right]$ with $v_{l,s,c}(z)$ the vacuum expectation value, and $\pi^{a}$ is the pseudoscalar field. Finally, the dilaton field $\phi$ in Eq. \eqref{action} only depends on the fifth dimension $z$ and explicit form is shown later in this section.

The equations of motion for each field can be obtained from varying the action in Eq. \eqref{action} with respect to the corresponding field. In order to find the vacuum expectation value,  one needs to remove all the fields and  keep only the background. The zeroth order of the action for the background field is given by

\begin{equation}
\begin{aligned}
S^{(0)}=&- \frac{1}{4} \int_{\epsilon}^{z_{m}} d^{5} x\left\{\frac{e^{-\phi(z)}}{z^{3}}\left(2 v_{l}^{\prime}(z) v_{l}^{\prime}(z)+v_{s}^{\prime}(z) v_{s}^{\prime}(z)+v_{c}^{\prime}(z) v_{c}^{\prime}(z)\right) - \right.\\
&\left.\frac{e^{-\phi(z)}}{z^{5}} \left(3 \left(2 v_{l}(z)^{2}+v_{s}(z)^{2}+v_{c}(z)^{2}\right)- \frac{\kappa}{4} \left(2 v_{l}(z)^{4}+v_{s}(z)^{4}+v_{c}(z)^{4}\right) \right)\right\}.
\end{aligned}
\label{zeroorder}
\end{equation}

The equation of motion for the scalar vacuum expectation value $v_{l,s,c}(z)$ is obtained as 

\begin{equation}
- \frac{z^3}{e^{-\phi}} \partial_{z} \frac{e^{-\phi}}{z^3} \partial_{z} v_{q}(z) - \frac{3}{z^2} v_{q}(z) - \frac{\kappa}{2 z^2} v_{q}^{3}(z)=0 ,
\label{VZEOM}
\end{equation}
where $q=l,s,c$. The solution for the scalar vacuum expectation value $v_{l,s,c}(z)$ that preserves the UV and IR asymptotic behavior is provided and justified in Ref.  \cite{Gherghetta:2009ac}

\begin{equation}  
 v(z)=a z + b z \tanh \left(c z^2\right),
 \label{vz}
\end{equation}
with the definitions for the parameters $a$, $b$, and $c$ as 
$$
a=\frac{\sqrt{3} m_q}{g_5 }, \quad b=\sqrt{\frac{4 \mu^{2}}{\kappa}}-a, \quad c=\frac{g_5 \sigma}{\sqrt{3} b},
$$
where $m_{q}$ is the quark mass and $\sigma$ is the chiral condensate. It worth noting that the UV and IR asymptotic behaviour of the $v(z)$ can be achieved by expanding Eq. \eqref{vz} at small and large $z$ as

\begin{equation}
  v(z \to 0) = a z +  b c z^{3} + \mathcal{O}(z^{5}), 
\end{equation}

\begin{equation}
  v(z \to \infty) = (a +  b) z = \sqrt{\frac{4 \mu^{2}}{\kappa}} z.
\end{equation}

In the initial soft wall model \cite{Karch:2006pv}, the dilaton field was originally characterized by the expression $\phi(z\to \infty) = \mu^{2} z^{2}$. Here, the parameter $\mu$ is connected to the Regge slope, establishing the mass scale for the meson spectrum and ensuring the presence of linear mass trajectories. Moreover, one can find the dilaton profile by substituting the equation \eqref{vz} into equation \eqref{VZEOM} and solve for $\phi$ field \cite{Gherghetta:2009ac}. However, in this approach the profile of the dilaton field exhibits dependence on the quark flavor and differs for each value of $v_{q}$. While this flavor reliance of the dilaton field poses no issue when exclusively considering light quarks, it becomes evident and inevitable when addressing heavy quarks like the charm quark \cite{Ahmed:2023zkk}. In Ref. \cite{Chelabi:2015cwn} a modified dilaton profile proposed with a negative quadratic dilaton at UV and a positive quadratic dilaton at IR which is different from the one obtained in Ref. \cite{Gherghetta:2009ac,Ahmed:2023zkk}, where positive quadratic dilaton is required at both UV and IR. In our present study, focusing solely on the IR asymptotic behavior of the $\phi$ field suffices for the numerical computations, thereby obviating the need to address the flavor-related variability of the dilaton profile.

The equation of motion for the vector, axial vector, and pseudoscalar mesons can be obtained from the expansion of the action in Eq. \eqref{action} up to the second order,

\begin{equation}
\begin{aligned}
S^{(2)}=&-\int d^{5} x\left\{\eta^{M N} \frac{e^{-\phi(z)}}{z^{3}}\left(\left(\partial_{M} \pi^{a}-A_{M}^{a}\right)\left(\partial_{N} \pi^{b}-A_{N}^{b}\right) M_{A}^{a b}-V_{M}^{a} V_{N}^{b} M_{V}^{a b}\right)\right.\\
&\left.+\frac{e^{-\phi(z)}}{4 g_{5}^{2} z} \eta^{M P} \eta^{N Q}\left(V^{a}_{M N} V^{b}_{P Q}+A^{a}_{M N} A^{b}_{P Q}\right)\right\},
\end{aligned}
\label{S2}
\end{equation}
where $\eta^{M N}$ is the metric in 5-D Minkowski space, $V^{a}(A^{a})_{M N}= \partial_{M} V^{a}(A^{a})_{N} - \partial_{N} V^{a}(A^{a})_{M}$. The mass terms in the action $M_{A}^{a b}$ and $M_{V}^{a b}$ are defined by

\begin{equation}
\begin{aligned}
&M_{A}^{a b} \delta^{a b}= Tr\left( \{t^{a},X_{0}\} \{t^{b},X_{0}\} \right), \\
&M_{V}^{a b} \delta^{a b}= Tr\left( [t^{a},X_{0}] [t^{b},X_{0}] \right), \\
\end{aligned}
\end{equation}
where $M_{V}^{a b}$ is zero for $a,b= 1,2,3,8,15$. The vector field in Eq. \eqref{S2} satisfies the following equation of motion,

\begin{equation}
- \partial^{M} \frac{e^{-\phi}}{g_{5}^{2} z} V_{M N}^{a} - \frac{e^{-\phi}}{ z^{3}} \left( M_{V}^{a a} V_{M}^{a} \right)=0 .
\label{emVm}
\end{equation}

The gauge choice for the vector field is set to $V_{z}^{a}=0$ and $\partial^{\mu} V_{\mu \perp}^{a}=0$ where $ V_{\mu \perp}^{a}$ is the transverse part of the vector field $ V_{\mu}^{a} = V_{\mu \perp}^{a} + V_{\mu \parallel}^{a}$. Considering the gauge fixing and then applying the 4D Fourier transformation, Eq. \eqref{emVm} reduces to the following

\begin{equation}
\left(-\frac{z}{e^{-\phi}} \partial_{z} \frac{e^{-\phi}}{z} \partial_{z}-\frac{2 g_{5}^{2} M_{V}^{a a}}{z^{2}}\right) V_{\mu \perp}^{a}(q, z)=-q^{2} V_{\mu \perp}^{a}(q, z).
\label{eqV}
\end{equation}
with $V_{\mu \perp}^{a}(q, z)$ is the 4D Fourier transformation of $V_{\mu \perp}^{a}(x, z)$. According to the AdS/CFT principle, it is allowed to write the transverse part of the vector field in terms of the bulk-to-boundary propagator and its boundary value at the UV regime, which acts as a Fourier transformation of the source of the 4D conserved vector current operator, $V_{\mu \perp}^{a}(q, z)= V_{\mu \perp}^{0a}(q) \mathcal{V}^{a}(q^{2},z)$. The boundary conditions for the bulk-to-boundary propagator $\mathcal{V}^{0a}(q^{2},z)$ to satisfies the equation of motion \eqref{eqV} are $\mathcal{V}^{a}(q^{2},\epsilon)=1$ and $\partial_{z} \mathcal{V}^{a}(q^{2},z_m)=0$. Moreover, the bulk-to-boundary propagator can be written as a sum over the meson poles

\begin{equation}
    \mathcal{V}^{a}(q^{2},z)= \sum_{n} \frac{- g_{5} f_{V^{n}}^{a} \psi_{V^{n}}^{a}(z)}{q^{2} - m_{V^{n}}^{a^{2}}},
\label{bulktoboundaryV}
\end{equation}
where $\psi_{V^{n}}(z)$ is a wavefunction which satisfies Eq. \eqref{eqV} with the boundary conditions $\psi_{V^{n}}(\epsilon)=0$ and $\partial_{z} \psi_{V^{n}}(z_m)=0$, and normalized as $ \int dz \frac{e^{-\phi}}{z} \psi^{n}_{V}(z) \psi^{m}_{V}(z)=\delta^{n m}$, and $f_{V^{n}}^{a}= |\partial_{z} \psi_{V^{n}}^{a}(\epsilon)/(g_{5} \epsilon) |$ is the decay constant of the $n^{th}$ mode of the vector meson \cite{Erlich:2005qh}.

Similar to the vector field, The axial vector field $A_{\mu}^{a}$ can be decomposed to the transverse and longitudinal parts, $A_{\mu}^{a}= A_{\mu \perp}^{a} + A_{\mu \parallel}^{a}$, where the longitudinal part $A_{\mu \parallel}^{a}=\partial_{\mu}\phi^{a}$ has the contribution to the pesudoscalar mesons. The equation of motion derived from Eq. \eqref{action} is given by

\begin{equation}
\left(-\frac{z}{e^{-\phi}} \partial_{z} \frac{e^{-\phi}}{z} \partial_{z}+\frac{2 g_{5}^{2}M_{A}^{a a}}{z^{2}}\right) A_{\mu \perp}^{a}(q, z)=-q^{2} A_{\mu \perp}^{a}(q, z),
\label{eqA}
\end{equation}
with the conditions $A_{z}^{a}=0$, and $\partial^{\mu} A_{\mu \perp}^{a}$, respectively. The bulk-to-boundary propagator of the axial vector field $\mathcal{A}^{a}(q^{2},z)$ satisfy the boundary conditions $\mathcal{A}^{a}(q^{2},\epsilon)=0$ and $\partial_{z} \mathcal{A}^{a}(q^{2},z_m)=0$, in the UV and IR region, also can be written as

\begin{equation}
    \mathcal{A}^{a}(q^{2},z)= \sum_{n} \frac{- g_{5} f_{A^{n}}^{a} \psi_{A^{n}}^{a}(z)}{q^{2} - m_{A^{n}}^{a^{2}}},
\label{bulktoboundaryA}
\end{equation}
with the wavefunction $\psi_{A^{n}}^{a}(z)$, and decay constant of the axial vector mesons $f_{A^{n}}^{a}= |\partial_{z} \psi_{A^{n}}^{a}(\epsilon)/(g_{5} \epsilon) |$.

And last but not least, the mass spectra of the pseudoscalar mesons can be obtained by solving the coupled equation of motions between the pseudoscalar field $\pi$ and the longitudinal part of the axial vector field $\phi$,
\begin{equation}
\begin{aligned}
& q^2 \partial_z \varphi^a(q, z)+\frac{2 g_5^2 M_A^{a a}}{z^2} \partial_z \pi^a(q, z)=0, \\
& \frac{z}{e^{-\phi}} \partial_z\left(\frac{e^{-\phi}}{z} \partial_z \varphi^a(q, z)\right)-\frac{2 g_5^2 M_A^{a a}}{z^2}\left(\varphi^a(q, z)-\pi^a(q, z)\right)=0,
\end{aligned}
\label{empesudo}
\end{equation}
with the boundary conditions $\pi^{a}(q^{2},\epsilon)=\phi^{a}(q^{2},\epsilon)=0$ and $\partial_{z}\pi^{a}(q^{2},z_m)=\partial_{z}\phi^{a}(q^{2},z_m)=0$. The bulk-to-boundary propagator for the longitudinal part of the axial vector field $\phi(q^{2},z)$ and pseudoscalar field $\pi (q^{2},z)$ are written as
\begin{equation}
\begin{aligned}
&\mathcal{\phi}(q^{2},z)= \sum_{n} \frac{ g_{5} m_{\pi^{n}}^{2} f_{\pi^{n}} \phi^{n}(z)}{q^{2} - m_{\pi^{n}}^{2}}, \\
&\mathcal{\pi}(q^{2},z)= \sum_{n} \frac{ g_{5} m_{\pi^{n}}^{2} f_{\pi^{n}} \pi^{n}(z)}{q^{2} - m_{\pi^{n}}^{2}},
\end{aligned}
\end{equation}
where $f_{\pi^{n}}= |\partial_{z} \phi^{n}(\epsilon)/(g_{5} \epsilon) |$ is the decay constant of the $n^{th}$ mode of the psuedoscalar meson.

\section{Three-point interactions and semileptonic form factors}
\label{semi-form}

\begin{figure}[ht!]
  \centering
  \includegraphics[width=0.6\linewidth]{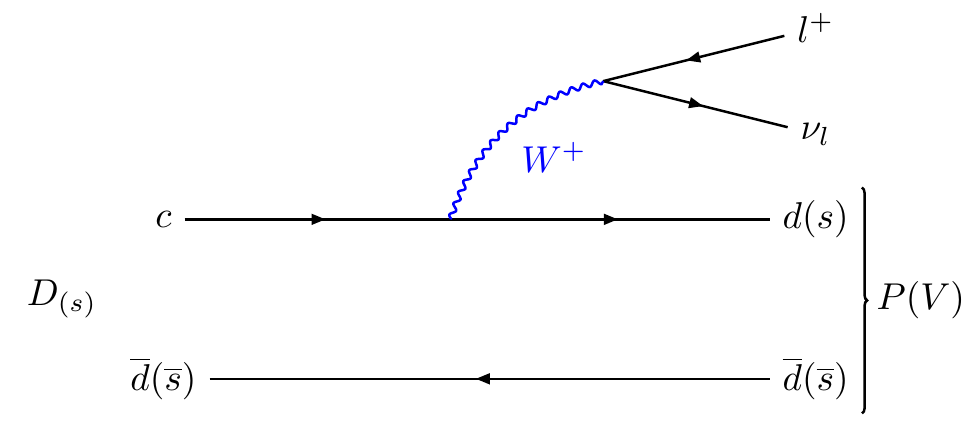}   
 
\caption{\footnotesize Feynman diagram for the semileptonic decay of $D_{(s)}$ into a pesudoscalar P (vector V) and $l^{+} \nu_{l}$.  }
\label{feynman}
\end{figure}

In this section, the semileptonic form factors of $D_{(s)} \to (P, V) l^{+} \nu_{l}$ are derived in the soft-wall holographic model. The Feynman diagram of the semileptonic decay process of $D_{(s)}$ to a pseudoscalar or a vector meson is shown in Fig. \ref{feynman}, where the charm quark goes through the process of $c \to d(s) W^{+} \to d(s) l^{+} \nu_{l}$. The matrix elements of the semileptonic decays of the $D_{(s)}$ meson within the SM is defined by \cite{Ivanov:2019nqd}

\begin{equation}
    \mathcal{M}\left(D_{(s)} \to (P,V) l^{+} \nu_{l}   \right)= \frac{G_{F}}{\sqrt{2}} V_{cq}^{*} \left< (P,V)|\bar{q}\gamma^{\mu}(1- \gamma_{5})c | D_{(s)}\right> \bar{\nu}_{l} \gamma^{\mu}(1- \gamma_{5})l,
\end{equation}
where $G_{F}$ is a fermi constant, $V_{cq}^{*}$ elements of a CKM matrix, and the hadronic and leptonic currents are given by the terms $\left< (P,V)|\bar{q}\gamma^{\mu}(1- \gamma_{5})c | D_{(s)}\right> $ and $ \bar{\nu}_{l} \gamma^{\mu}(1- \gamma_{5})l$, respectively. The hadronic current can be parameterized in terms of the invariant form factors, which depend on the momentum transfer squared ($q^{2}$). For the case of the pseudoscalar mesons in the final state, only the vector current ($\bar{q}\gamma^{\mu}c$) contributes to the form factors. The transition form factors are defined by \cite{Wirbel:1985ji}

\begin{equation}
\begin{aligned}
\left\langle P\left(p_2\right)\left|V^\mu\right| D_{(s)}\left(p_1\right)\right\rangle= & F_+\left(q^2\right)\left[P^\mu-\frac{M_1^2-M_2^2}{q^2} q^\mu\right]+F_0\left(q^2\right) \frac{M_1^2-M_2^2}{q^2} q^\mu \\
\left\langle V\left(p_2, \epsilon_2\right)\left|V^\mu-A^\mu\right| D_{(s)}\left(p_1\right)\right\rangle= & -\left(M_1+M_2\right) \epsilon_2^{* \mu} A_1\left(q^2\right)+\frac{\epsilon_2^* \cdot q}{M_1+M_2} P^\mu A_2\left(q^2\right) \\
& +2 M_2 \frac{\epsilon_2^* \cdot q}{q^2} q^\mu\left[A_3\left(q^2\right)-A_0\left(q^2\right)\right]+\frac{2 i \varepsilon_{\mu \nu \rho \sigma} \epsilon_2^{* \nu} p_1^{\rho} p_2^{\sigma}}{M_1+M_2} V\left(q^2\right),
\end{aligned}
\end{equation}
where $P=p_{1}+p_{2}$, $q=p_{1}-p_{2}$, $M_{1}$ and $M_{2}$ are the mass of the mesons in the initial and final state, respectively, and $\epsilon_{2}$ is the polarization vector of the final vector meson. The $A_{3}(q^{2})$ form factor is not independent and can be written as a combination between $A_{1}(q^{2})$ and $A_{2}(q^{2})$. For the present study, we only consider the form factors associated with the vector meson exchange $F_{+}(q^{2})$ and $V(q^{2})$, and axial vector meson exchange $A_{1}(q^{2})$, since these are the most important form factors in the limit of zero lepton mass. 

Using the holographic QCD approach, the semileptonic form factors can be deduced from the three-point functions \cite{Abidin:2009aj,Momeni:2022gqb}. The cubic terms of the 5D action used to find the $F_{+}(q^{2})$ are $S(V\pi\pi)$, and for $V(q^{2})$ and $A_{1}(q^{2})$ are $S(VV\pi)$ and $S(VA\pi)$, respectively. The expansion of the 5D action \eqref{action} to cubic order is given by,

\begin{equation}
\begin{aligned}
S^{(3)}= & -\int d^{5} x\left\{\eta^{M N} \frac{e^{-\phi(z)}}{z^{3}}(2\left(A_{M}^{a}-\partial_{M} \pi^{a}\right) V_{N}^{b} \pi^{c} g^{a b c}+V_{M}^{a}\left(\partial_{N}\left(\pi^{b} \pi^{c}\right)-2 A_{M}^{b} \pi^{c}\right) h^{a b c}\right.\\
& -  V_{M}^{a} V_{N}^{b} \pi^{c} k^{a b c}) +\frac{e^{-\phi(z)}}{2 g_{5}^{2} z} \eta^{M P} \eta^{N Q}(V_{M N}^{a} V_{P}^{b} V_{Q}^{c}+V_{M N}^{a} A_{P}^{b} A_{Q}^{c}+A_{M N}^{a} V_{P}^{b} A_{Q}^{c}
.\\ 
&\left. +A_{M N}^{a} A_{P}^{b} V_{Q}^{c}) f^{b c a}\right\}
\end{aligned}
\end{equation}
with the following definitions for $g^{a b c}$, $h^{a b c}$, and $k^{a b c}$,

\begin{equation}
\begin{aligned}
&g^{a b c} = i Tr\left( \{t^{a},X_{0}\} [t^{b},\{t^{c},X_{0}\}] \right), \\
&h^{a b c} = i Tr\left( [t^{a},X_{0}] \{t^{b},\{t^{c},X_{0}\}\} \right), \\
&k^{a b c} = -2 Tr\left( [t^{a},X_{0}] [t^{b},\{t^{c},X_{0}\}] \right). \\
\end{aligned}
\end{equation}

In the present work, we are interested in the three-point interactions of the $V\pi\pi$, $VV\pi$, and $VA\pi$. The corresponding part of the action to these three-point interactions are

\begin{equation}
\begin{aligned}
S_{V\pi\pi}= & -\int_{\epsilon}^{z_{m}} d^{5} x\left\{\eta^{M N} \frac{e^{-\phi(z)}}{z^{3}}\left(2\left(A_{M}^{a}-\partial_{M} \pi^{a}\right) V_{N}^{b} \pi^{c} g^{a b c}+V_{M}^{a}\left(\partial_{N}\left(\pi^{b} \pi^{c}\right)-2 A_{N}^{b} \pi^{c}\right) h^{a b c}\right)\right.\\
&\left. +\frac{e^{-\phi(z)}}{2 g_{5}^{2} z} \eta^{M P} \eta^{n Q}\left(V_{M N}^{a} A_{P}^{b} A_{Q}^{c}\right) f^{abc}\right\}
\end{aligned}
\label{actionpion}
\end{equation}

\begin{equation}
    S_{VV\pi}=  \int d^{5} x\frac{e^{-\phi(z)}}{ z^{3}} \eta^{M N} \left(V_{M}^{a} V_{N}^{b} \pi^{c}\right) k^{a b c}
\label{VVpi}
\end{equation}

\begin{equation}
    S_{VA\pi}= -\int_{\epsilon}^{z_{m}} d^{5} x\left\{2 \eta^{M N} \frac{e^{-\phi(z)}}{z^{3}} A_{M}^{a} V_{N}^{b} \pi^{c} \left( g^{a b c} -  h^{b a c}\right) +\frac{e^{-\phi(z)}}{2 g_{5}^{2} z} \eta^{M P} \eta^{N Q}\left(V_{M N}^{a} A_{P}^{b} A_{Q}^{c}\right) f^{abc}\right\}
\label{VApi}
\end{equation}

Similar to the derivation of the electromagnetic form factors using the three point function \cite{Ahmed:2023zkk}, and semileptonic form factors in the work of Refs. \cite{Abidin:2009aj,Momeni:2022gqb}, one can obtain the $F_{+}(q^{2})$, $V(q^{2})$, and $A_{1}(q^{2})$ as the following,

\begin{equation}
   F_{+} (q^{2})=  \int dz  \frac{e^{-\phi(z)}}{ z} \left( f^{abc} \partial_{z}\phi^{a} \mathcal{V}^{b}(q^{2},z) \partial_{z}\phi^{c} - \frac{2 g_{5}^{2}}{z^{2}} (\pi^{a}-\phi^{a}) \mathcal{V}^{b}(q^{2},z) (\pi^{c}-\phi^{c})  (g^{abc} - h^{bac}) \right) ,
\label{fplus}
\end{equation}

\begin{equation}
    V(q^{2})=\frac{(M_{1}+M_{2}) g_{5}^{2}}{2} \int dz  \frac{e^{-\phi(z)} }{ z^3} k^{abc}  V^{a}(z) \mathcal{V}^{b}(q^{2},z) \pi^{c}(z),
\label{Vform}
\end{equation}

\begin{equation}
\begin{aligned}
    A_{1}(q^{2})= & \int dz \frac{e^{-\phi(z)}}{ z}  \left(\frac{M_{1}^{2} + M_{2}^{2} - q^{2}}{2 (M_{1}+M_{2})} \right)   f^{bac} \mathcal{A}^{a}(q^{2},z) V^{b}(z) \phi^{c}(z)\\ 
& - \int dz \frac{e^{-\phi(z)}}{ z^{3}}  \frac{2 g_{5}^{2} }{(M_{1}+M_{2})}  \mathcal{A}^{a}(q^{2},z) V^{b}(z) \pi^{c}(z) (g^{abc} - h^{bac})  .
\end{aligned}
\label{A1form}
\end{equation}

\section{Results}
\label{result}
In this section, we show the numerical results for the meson masses and decay constants of the vector, axial vector, and pseudoscalar mesons at the ground state and the form factors of the semileptonic decay process of $D_{(s)}$ mesons to a pseudoscalar or vector mesons within the framework of $N_{f}=4$ holographic QCD.

Firstly, let us set the parameters of the model. The parameters of the model that can be found from the fitting to the experimental data are $\mu$, $m_{u}$, $m_{s}$, $m_{c}$, $\sigma_u$, $\sigma_s$, $\sigma_c$, $\kappa$ and $z_{m}$. The value of $\mu$ is found to be $430$ MeV from the fitting of the experimental masses of the ground and higher excited states of the $\rho$ meson. Since the pion decay constant and pion mass are related to the light quark mass and condensate by the Gell-Mann-Oakes-Renner (GOR) relation, $f_{\pi}^{2} m_{\pi}^{2} = 2 m_{q} \sigma$, the measured value of the pion decay constant $f_{\pi}=92.4$ MeV and pion mass $m_{\pi}=139.6$ MeV were used to adjust the up quark mass and up quark condensate. Similarly, we use the GOR relation to fix the values of $m_{s}$ and $\sigma_s$ from  the measured mass and decay constant of the kaon. After fixing $\mu$, $m_{u}$, and $\sigma_u$, one can use the experimental value of the $a_{1}$ meson to determine the value of $\kappa$. For the parameters of the charm sector, the mass $m_{c}$ and charm quark condensate $\sigma_c$ are found from the fitting of the model with the experimental value of the masses $m_{\eta_{c}}$ and $m_{\chi_{c1}}$. Following the work of Refs. \cite{Ahmed:2023zkk,Chen:2021wzj}, the value of $z_{m}$ is fixed at $10$ GeV. The numerical values of the parameters are provided in Table \ref{tab:freepar}. 

By using the parameters in Table \ref{tab:freepar}, one can obtain the ground state mass and decay constants of the vector, axial vector, and pseudoscalar mesons. Table \ref{tab:massanddecay} presents the results of the masses and decay constants. It is worth noting that the $SU(4)$ flavor symmetry is explicitly breaking due to the different values of the quark masses and condensates. And the consequence of the flavor symmetry breaking is the difference between the masses of the strange and charmed mesons with the light flavor mesons. However, in the vector sector the mass $M_{V}^{a a}$ in Eq. \eqref{eqV} is zero for $a= 1, 2, 3, 8, 15$, and this returns the same masses for the $\rho$, $\omega$, and $J/\Psi$ mesons. This issue solved for the $J/\Psi$ meson by adding an auxiliary heavy field to the action, which only include the contribution of the charm quark to explicitly break the $SU(4)_{V}$ to $SU(3)_{V}$ \cite{Chen:2021wzj}. Since the contributions of $\omega$ and $J/\Psi$ mesons are not important for scope of the current work, we did not include the auxiliary field in the 5D action.

\begin{table}
\center
\begin{tabular}{c  c  c c c}
\hline
\hline

    $m_u = 3.2$ &    &   $\sigma_u = (296.2)^3$ &   &   $\mu = 430$   \\ 
    $m_s =142.3$  &  &   $\sigma_s = (259.8)^3$ &   &  $\kappa =30$   \\ 
    $m_c =1597.1$ &  &   $\sigma_c = (302)^3$ &      & $z_{m}=10000$  \\ 
     \hline
     \hline
\end{tabular}
\caption{\footnotesize The values of the free parameters with the unit of MeV.}
\label{tab:freepar}
\end{table}

\begin{table}
\center
%\begin{adjustbox}{width=\textwidth}
 \begin{tabular}{c c c c c c c c c}
\hline
\hline
          Meson   &    & Mass (MeV) &     & Measured (MeV) &    & Decay constant (MeV) &    & Measured (MeV)   \\ 
\hline

        $\rho$    &    & 860        &     &    775         &    & 288.5                &    & 345              \\
        $K^{*}$   &    & 860.1      &     &    892         &    & 288.3                &    &                  \\
        $D^{*}$   &    & 1914.9     &     &    2007        &    & 413.4                &    &                  \\
      $D_{s}^{*}$ &    & 1911.4     &     &    2112        &    & 427.8                &    &                  \\
        $a_{1}$   &    & 1286.95    &     &    1230        &    & 351.3                &    & 354              \\
        $K_{1}$   &    & 1287.7     &     &    1253        &    & 348.3                &    &                  \\
        $f_{1}$   &    & 1287.97    &     &    1282        &    & 346.8                &    &                  \\
        $D_{1}$   &    & 2641.5     &     &    2422        &    & 502.11               &    &                  \\
       $D_{s1}$   &    & 2657.55    &     &    2460        &    & 475.74               &    &                  \\
        $\pi$     &    & 141.7      &     &    139.6       &    & 91.03                &    & 92.07            \\
        $K$       &    & 622.2      &     &    498         &    & 108.5                &    & 110              \\
        $\eta$    &    & 740.52     &     &    548         &    & 126.3                &    &                  \\
        $D^{0}$   &    & 2032.7     &     &    1865        &    & 199.3                &    & 149.8            \\
        $D_{s}$   &    & 2114.3     &     &    1968        &    & 197.7                &    & 176.1            \\
     
     \hline
     \hline
\end{tabular}
%\end{adjustbox}
\caption{\footnotesize The predicted masses and decay constants calculated from the hQCD compared to experimental or lattice data. The measured value of the mass of the vector, axial vector, and pseudoscalar mesons, and decay constant of the pseudoscalar mesons are taken from the particle data group (PDG) \cite{ParticleDataGroup:2020ssz}. The measured value of the decay constant of the $\rho$ and $a_{1}$ mesons are taken from Refs. \cite{Donoghue:1992dd} and \cite{Isgur:1988vm}, respectively. }
\label{tab:massanddecay}
\end{table}

Furthermore, we investigate the form factors of the following semileptonic decay processes, $D^{+} \to (\pi, K, \eta, K^{*}) l^{+} \nu_{l}$ and $D_{s}^{+} \to ( K, \eta, K^{*}) l^{+} \nu_{l}$. From the experimental point of view, the semileptonic decays are important to find the elements of the CKM matrix. For that reason, it is important to determine the maximum-recoil values of $F_{+}(q^{2}=0)$, and $V(q^{2}=0)$ and $A_{1}(q^{2}=0)$ for $D_{(s)}^{+} \to (\pi, K, \eta) l^{+} \nu_{l}$, and $D_{(s)}^{+} \to K^{*} l^{+} \nu_{l}$, respectively. Regarding the vector form factor for $D_{(s)}^{+} \to K^{*} l^{+} \nu_{l}$, it is more favorite to take the ratio between $V(q^{2}=0)$ and $A_{1}(q^{2}=0)$, $r_{v}=V(0)/A_{1}(0)$ \cite{BESIII:2018xre}. The comparison of the maximum-recoil values at $q^2 = 0$ with the experimental data, lattice QCD, and other theoretical approaches,e.g.,  light-cone sum rules (LCSR), light-front quark model (LFQM), constituent quark model(CQM), covariant confined quark model (CCQM)  are presented in Table \ref{tab:decay}. 

\begin{figure}[ht!]
  \centering
  \includegraphics[width=0.5\linewidth]{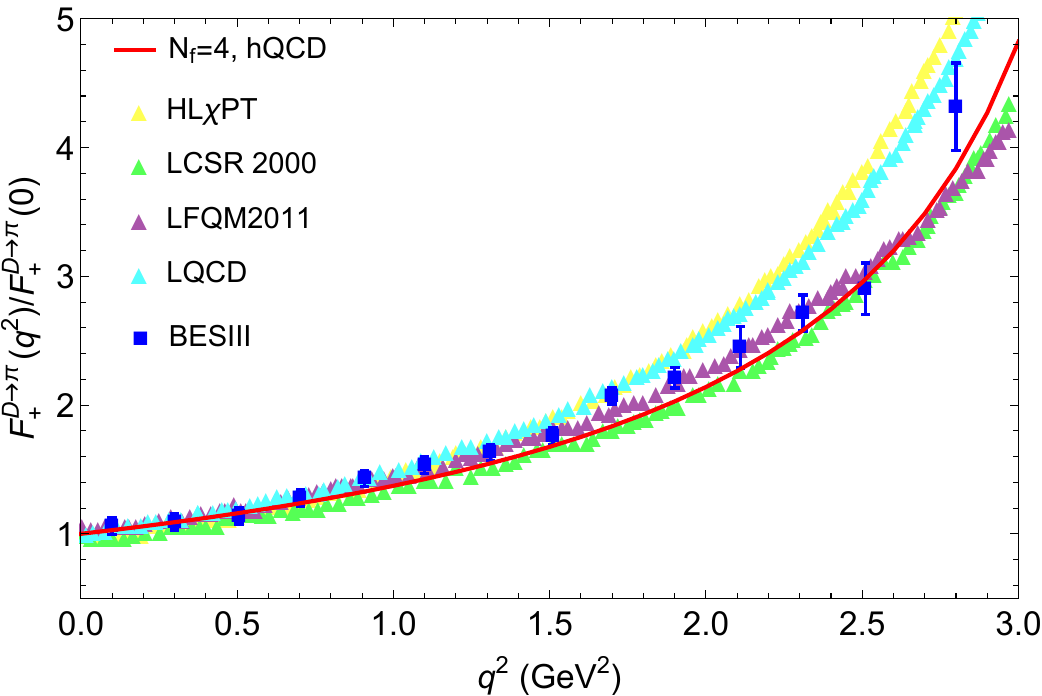}   

\caption{\footnotesize The semileptonic form factor $F_{+}(q^{2})$ for $D \to \pi l^{+} \nu_{l}$. Our result (solid red line) is compared with the experimental data (blue square)\cite{BESIII:2015tql}, lattice data (cyan triangle) \cite{FermilabLattice:2004ncd}, LFQM (purple triangle) \cite{Verma:2011yw}, LCSR (green triangle)\cite{Khodjamirian:2000ds}, and HL$\chi$PT (yellow triangle)\cite{Fajfer:2004mv}.  }
\label{formfpDpi}
\end{figure}

For the case of the pion in the final state, the form factor $f_{+}^{D\rightarrow \pi}(0)$ is consistent with the experimental data and lattice QCD with a small discrepancy of $6.75\%$ and $9\%$,respectively. Meanwhile to compare our full form factor with the others qualitatively, we normalize the form factors with the maximum-recoil values of $F_{+}(q^{2}=0)$. The result of the form factor for $D^{+} \to \pi l^{+} \nu_{l}$ is shown in Fig. \ref{formfpDpi}, where we compare our calculation with the experimental data \cite{BESIII:2015tql}, lattice QCD data \cite{FermilabLattice:2004ncd}, and different theoretical approaches like LCSR, LFQM, CQM, CCQM and heavy-light chiral perturbation theory (HL$\chi$PT) (See Ref. \cite{Palmer:2013yia} and the references therein). The result of $F_{+}(q^{2})$ is in excellent agreement with the experiment and Lattice QCD and has a better reproduction compared to other theoretical approaches. 

\begin{figure}[ht!]
  \centering
  \includegraphics[width=0.49\linewidth]{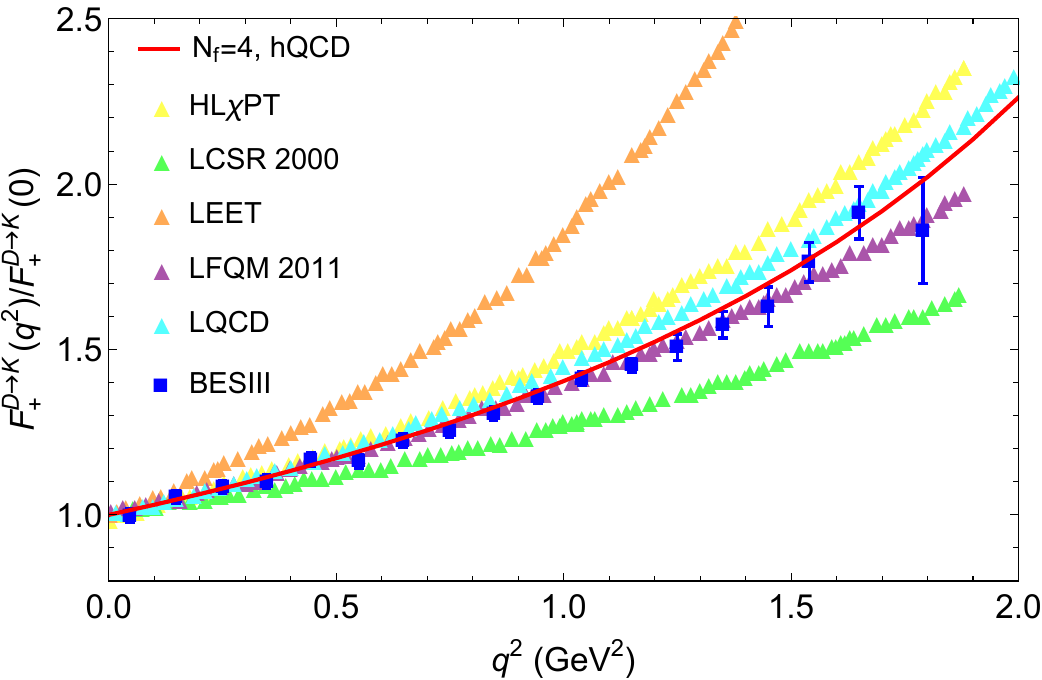}   
  \includegraphics[width=0.49\linewidth]{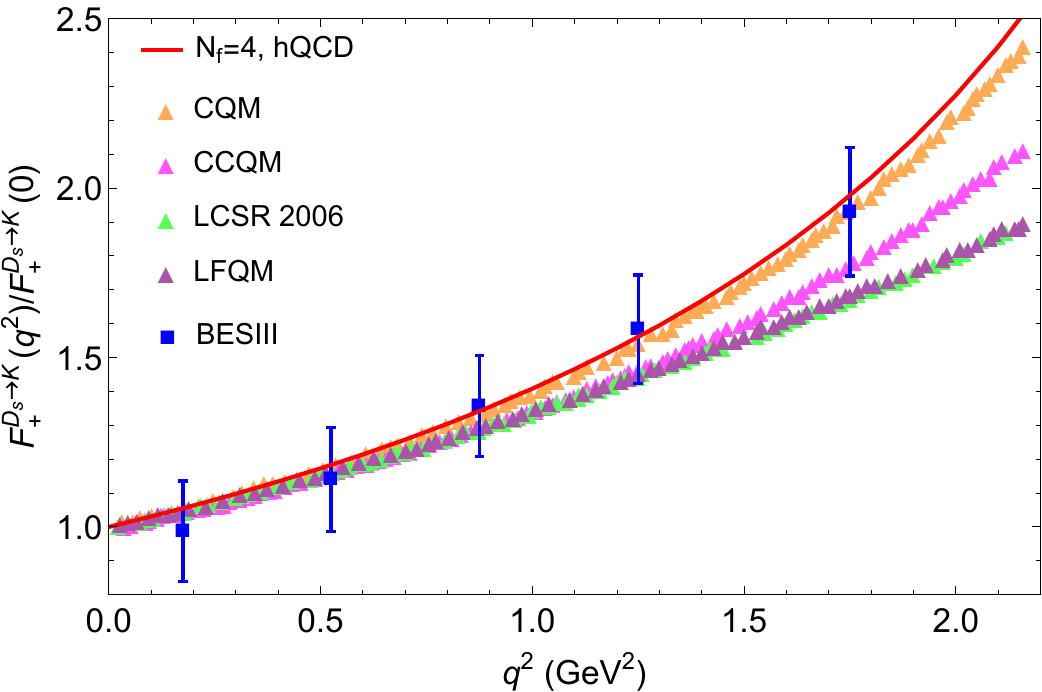}

\caption{\footnotesize Results of $F_{+}(q^{2})$ for the decay of $D_{(s)}$ to kaon. Left: our result for $D \to K l^{+} \nu_{l}$ (solid red line), the experimental data (blue square)\cite{BESIII:2015tql}, lattice data (cyan triangle) \cite{FermilabLattice:2004ncd}, LFQM (purple triangle) \cite{Verma:2011yw}, LEET (orange triangle) \cite{Charles:1998dr}, LCSR (green triangle)\cite{Khodjamirian:2000ds}, and HL$\chi$PT (yellow triangle)\cite{Fajfer:2004mv}. Right: $D_{s} \to K l^{+} \nu_{l}$ form factor (Solid red line) compared to the experimental data (Blue) \cite{BESIII:2018xre}, LFQM (purple) \cite{Verma:2011yw}, LCSR (green) \cite{Wu:2006rd}, CCQM (magenta) \cite{Soni:2018adu}, and CQM (orange) \cite{Melikhov:2000yu}. }
\label{formfpDK}
\end{figure}

In the case of $D_{(s)}\to K$, the form factor at zero momentum has more discrepancy compare to $D \to \pi$, which is $20\%$. This can be related to the fact that the mass of the Kaon is not well reproduced in the model as shown in Table \ref{tab:massanddecay}. However, as shown in Fig. \ref{formfpDK}, the normalized form factor $F_{+}(q^{2})$ aligns very well with the experimental and lattice QCD data and outperforming other theoretical approaches, such as LCSR, LFQM, CQM, CCQM, HL$\chi$PT and large energy effective theory (LEET) (see the caption of Fig. \ref{formfpDK} for the references). 

\begin{figure}[ht!]
  \centering
  \includegraphics[width=0.49\linewidth]{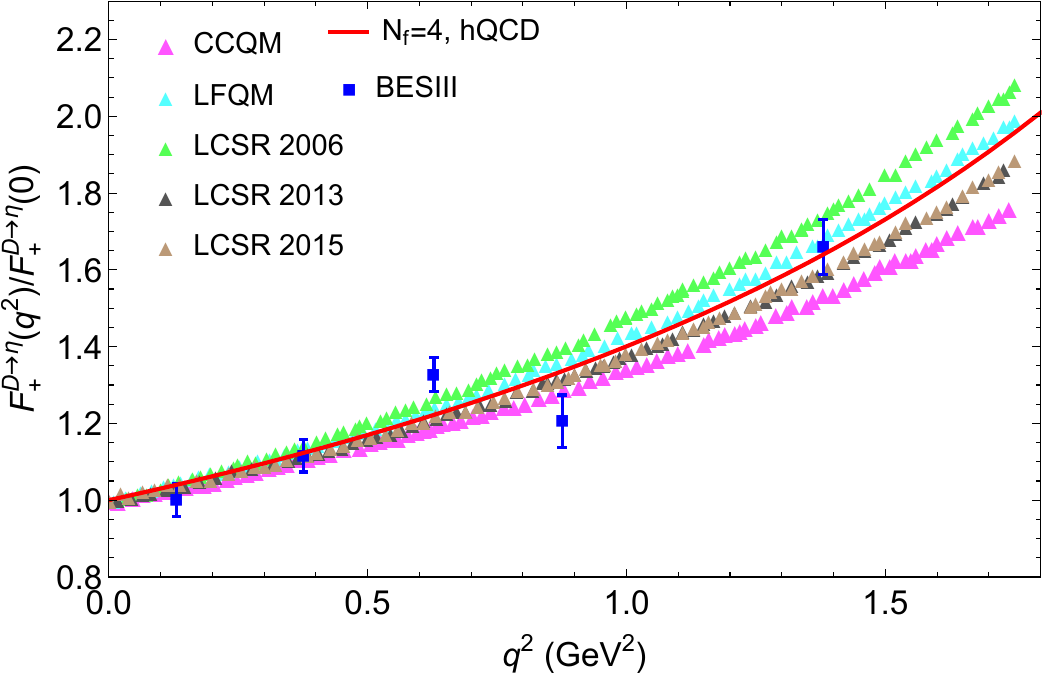} 
  \includegraphics[width=0.49\linewidth]{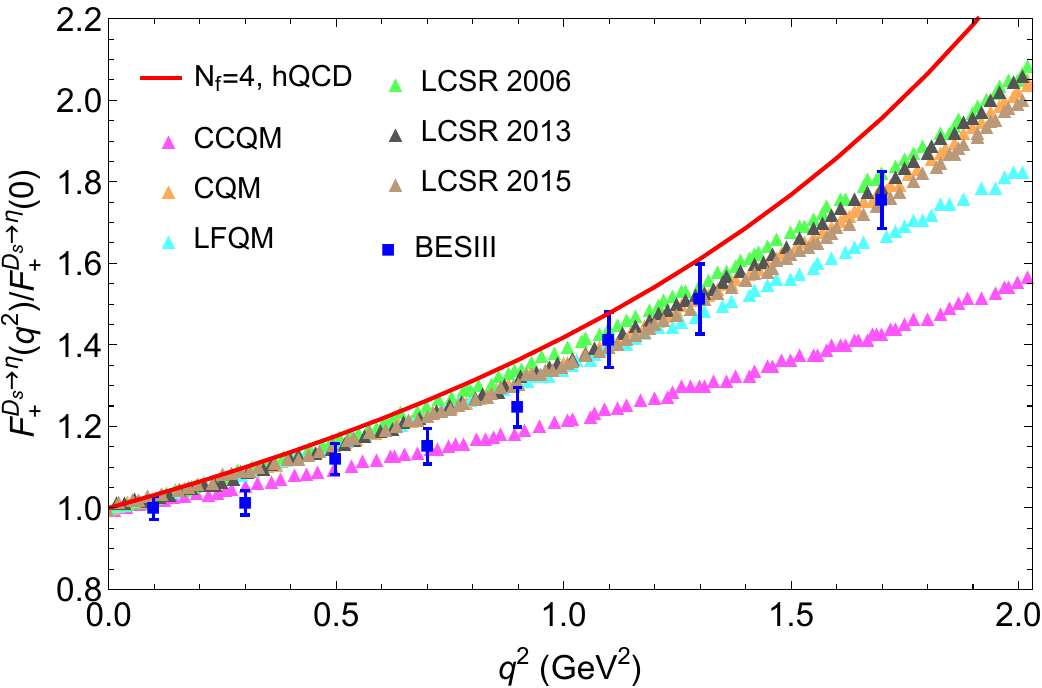}   
  
\caption{\footnotesize Form factor $F_{+}(q^{2})$ for $D_{(s)}^{+} \to  \eta$ in the present work (red), experimental data (blue) for $D$ \cite{Ablikim:2020hsc} and $D_{s}$ \cite{BESIII:2019qci}, CCQM (magenta) \cite{Soni:2018adu}, CQM (orange) \cite{Melikhov:2000yu}, LCSR (green) \cite{Wu:2006rd}, LCSR (black) \cite{Offen:2013nma}, and LCSR (brown) \cite{Duplancic:2015zna}.  }
\label{formfpDeta}
\end{figure}

The Experimental form factors of $D^{+} \to \eta^{(\prime)} l^{+} \nu_{l}$ are reported by the BESIII collaborations in Refs. \cite{BESIII:2019qci,Ablikim:2020hsc}. In the current analysis, we only study $D^{+} \to \eta l^{+} \nu_{l}$, and if one wants to consider the $\eta^{\prime}$ in the holographic QCD, the $U(1)_A$ axial anomaly should be considered \cite{Katz:2007tf}. The compatibility of the form factors of the $D^{+} \to  \eta $ decay with the experimental data \cite{Ablikim:2020hsc} and other theoretical frameworks can be seen in Fig. \ref{formfpDeta}. However, the result of the $D_{s}^{+} \to  \eta $ has some discrepancy with the experimental data \cite{BESIII:2019qci} and grows faster at large $q^{2}$. The discrepancy of the $D_{s}^{+} \to  \eta $ also can be seen from Table \ref{tab:decay} for $f_{+}^{D_{s} \rightarrow \eta}(0)$. It is worth noting that similar incompatibility with the experimental data has also been reported by other approaches such as LCSR, LFQM, CQM, CCQM and even lattice QCD has discrepancy with $25\%$.

\begin{figure}
  \centering
  \includegraphics[width=0.49\linewidth]{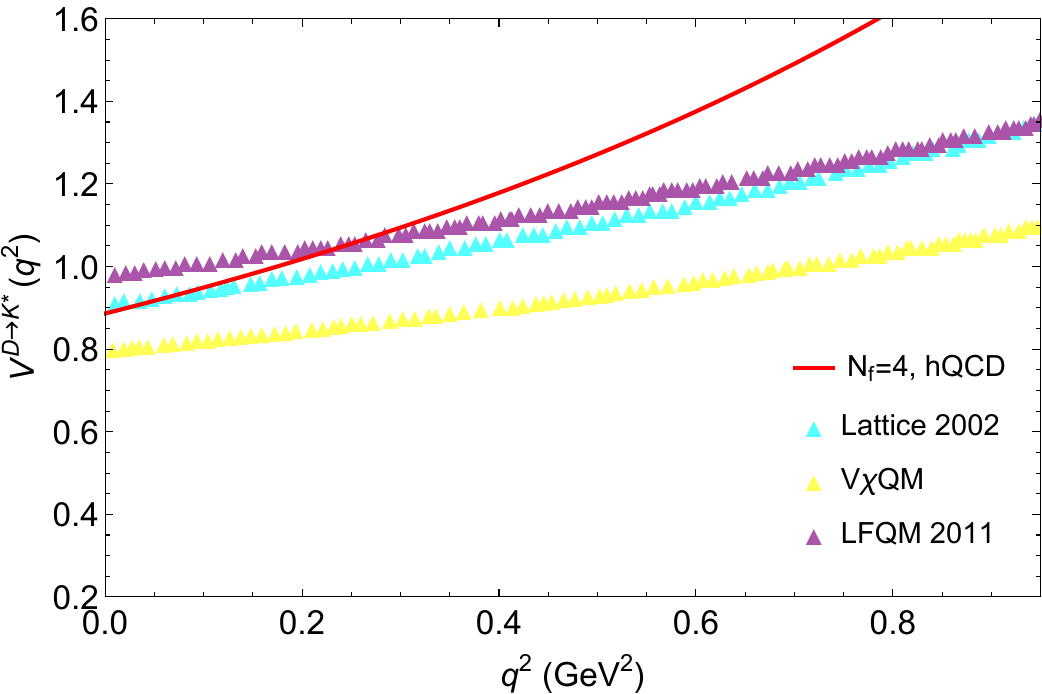}   
  \includegraphics[width=0.49\linewidth]{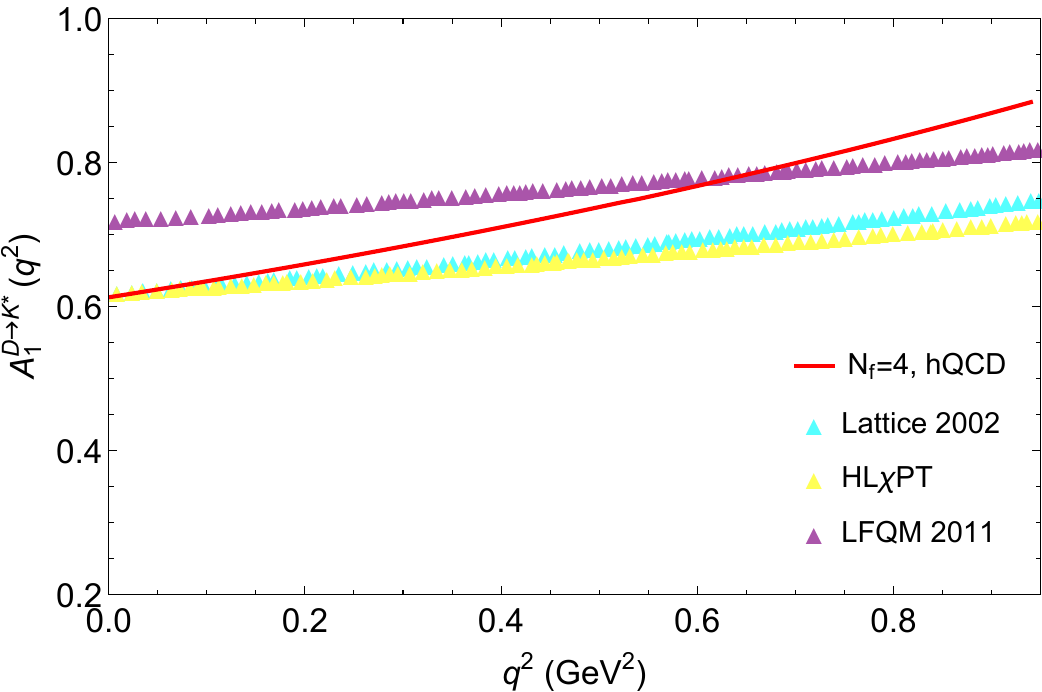}   
  
\caption{\footnotesize Comparison of the form factors $V(q^{2})$ (red) and $A_{1}(q^{2})$ (red) for $D \to K^{*}$ with different theoretical approaches. Lattice data (cyan) from Ref. \cite{Abada:2002ie}, LEV$\chi$QM (yellow) from Ref. \cite{Palmer:2013yia}, LFQM (purple) from Ref. \cite{Verma:2011yw}, and HL$\chi$PT (yellow) from Ref. \cite{Fajfer:2004mv}   }
\label{VDkstarD}
\end{figure}

\begin{figure}
  \centering
  \includegraphics[width=0.49\linewidth]{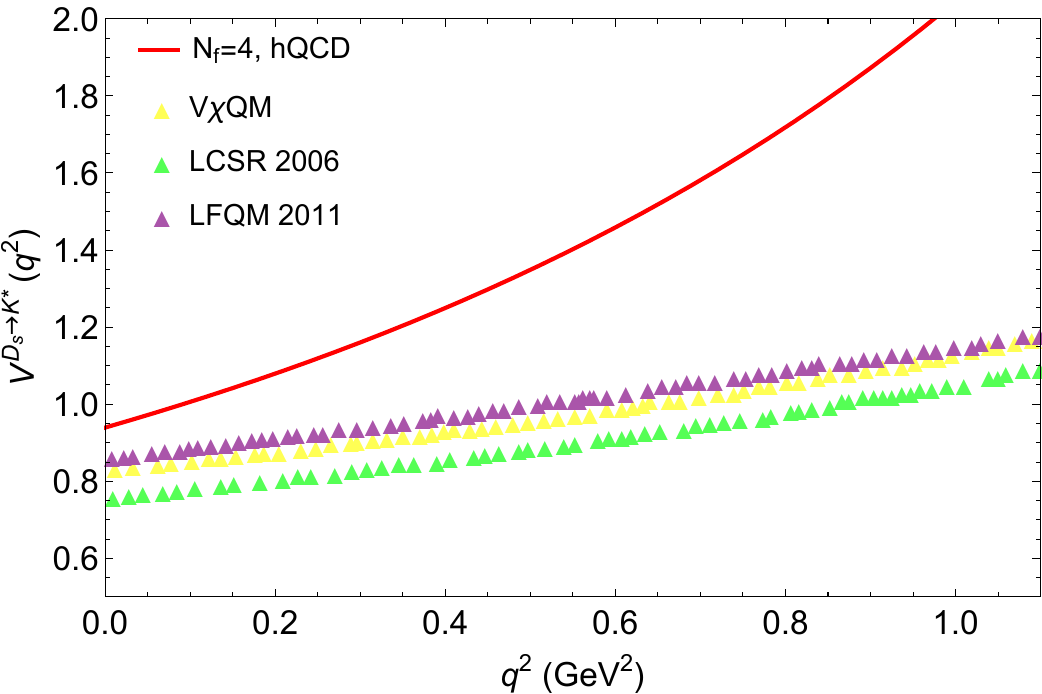}   
  \includegraphics[width=0.49\linewidth]{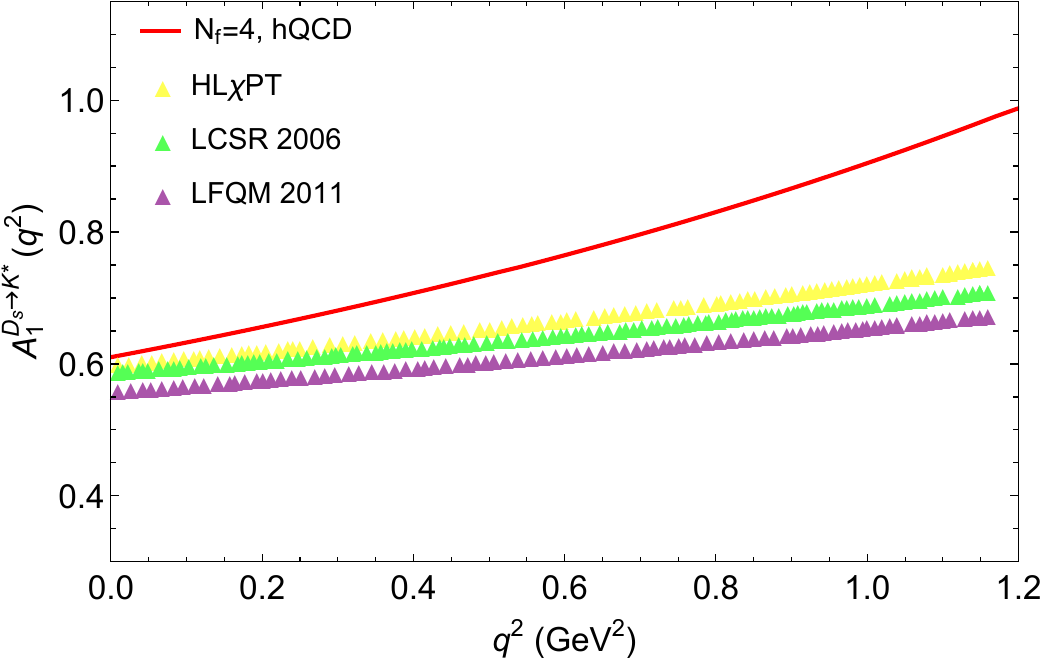}   
  
\caption{\footnotesize $D_{s} \to V$ form factors $V(q^{2})$ and $A_{1}(q^{2})$. The references for V$\chi$QM, LFQM, and HL$\chi$PT are similar to the one mentioned in Fig. \ref{VDkstarD}. LCSR is taken from Ref.   \cite{Wu:2006rd}.}
\label{VDkstarDs}
\end{figure}

Finally, we predict the vector form factors associated with the vector meson exchange $V(q^{2})$ and axial vector meson exchange $A_{1}(q^{2})$. As mentioned before, it is more interesting to compare their ratios at maximum recoil. From the experimental side, the form factors of $D \to K^{*}$ and $D_{s} \to K^{*}$ are not reported for the full range of momentum. However, Only the ratio of $r_{V}^{D\to K^{*}}$ and $r_{V}^{D_{s} \to K^{*}}$ are measured. Meanwhile the lattice QCD community calculated the $D \to K^{*}$ form factor. As shown in Table \ref{tab:decay}, our results of $r_{V}^{D\to K^{*}}$ and $r_{V}^{D_{s} \to K^{*}}$ are well aligned with the experimental data. The results of the $D \to K^{*}$ and $D_{s} \to K^{*}$ form factors are shown in Fig. \ref{VDkstarD} and Fig. \ref{VDkstarDs}, respectively. From  Fig. \ref{VDkstarD}, we can see that at the low value of $q^{2}$, our results are within the range of the other approaches and well consistent with lattice data \cite{Abada:2002ie}. However, by going to the high $q^{2}$, the form factors $V(q^{2})$ and $A_{1}(q^{2})$ are raised faster than other approaches. Similar feature can be seen for $D_{s} \to K^{*}$ as shown in Fig. \ref{VDkstarDs}, especially for the case of $V(q^{2})$. This can be regarded as a signal that, there maybe a missing information for the vector form factors associated with the vector meson exchange $V(q^{2})$ and axial vector meson exchange $A_{1}(q^{2})$ using the holographic QCD model.

\begin{table}
\center
%\begin{adjustbox}{width=\textwidth}
\begin{tabular}{ c  c c  c  c  c  c c  c  c  c  c c  c }
\hline
\hline
          FFs                          & hQCD  & LCSR \cite{Khodjamirian:2000ds} & LCSR \cite{Wu:2006rd} & LCSR \cite{Offen:2013nma} & LCSR \cite{Duplancic:2015zna}& LFQM \cite{Verma:2011yw} &  CQM \cite{Melikhov:2000yu} & CCQM \cite{Soni:2018adu} &   LQCD  & Exp.  \\ 
\hline
          $f_{+}^{D\rightarrow \pi}(0)$       &   0.58     &   0.65     &  0.635    &     -     &    -     &  0.66 & 0.69 & 0.63 &  0.64 \cite{FermilabLattice:2004ncd}  & 0.622  \cite{BESIII:2015tql} \\ 
\hline
          $f_{+}^{D\rightarrow K}(0)$         &   0.57     &   0.76     &  0.661    &     -     &    -     & 0.79  & 0.78 & 0.78 &  0.73 \cite{FermilabLattice:2004ncd}  & 0.725  \cite{BESIII:2015tql} \\ 
          
\hline      
          $f_{+}^{D_{s}\rightarrow K}(0)$     &   0.57     &    -       &  0.820    &     -     &    -     & 0.66  & 0.72 & 0.60 &  0.77 \cite{Lubicz:2017syv}  & 0.72  \cite{BESIII:2018xre} \\ 
\hline
          $f_{+}^{D\rightarrow \eta}(0)$      &   0.31     &    -       &  0.556    &    0.552  &  0.429   & 0.71  &  -   & 0.67 &    -    & 0.39  \cite{Ablikim:2020hsc} \\ 
\hline
        $f_{+}^{D_{s}\rightarrow \eta}(0)$    &   0.66     &    -       &  0.611    &    0.520  &  0.495   & 0.76  & 0.78 & 0.78 &  0.564 \cite{Bali:2014pva} & 0.45 \cite{BESIII:2019qci}  \\ 
\hline
          $r_{V}^{D\to K^{*}}$                &   1.40     &    -       &  1.385    &     -     &    -     & 1.36  & 1.56 & 1.22 &  1.468 \cite{Abada:2002ie} & 1.41 \cite{BESIII:2015hty}  \\ 
\hline
          $r_{V}^{D_{s}\to K^{*}}$            &   1.53     &    -       &  1.309    &     -     &    -     & 1.55  & 1.82 & 1.40 &     -   & 1.67 \cite{BESIII:2018xre}  \\
                                          
     \hline
     \hline
\end{tabular}
%\end{adjustbox}
\caption{\footnotesize  Comparison of the maximum-recoil values of the form factors with the different theoretical approaches, lattice QCD, and experimental data.}
\label{tab:decay}
\end{table}

\section{Conclusions}
\label{conclusion}

In this study, we utilized a modified soft-wall holographic model with four flavors to comprehensively investigate various aspects of mesons, including their spectra, decay constants, and semileptonic form factors. By fitting the model parameters to experimental meson masses, we successfully determined the mass and decay constants of vector mesons ($\rho$, $K^*$, $D^*$, and $D_s^*$), axial vector mesons ($a_1$, $K_1$, $f_1$, $D_1$, and $D_{s1}$), and pseudoscalar mesons ($\pi$, $K$, $\eta$, $D$, $D_s$). In the vector sector, we calculated the decay constants of $K^{*}$, $D^*$, and $D_s^*$ mesons and compared the result for the $\rho$ meson with experimental data, revealing a discrepancy of approximately $16\%$. However, in the axial vector sector, the decay constant of the $a_1$ meson exhibited excellent agreement with experimental data. Moreover, we successfully reproduced the decay constants of the pion and kaon in our model, comparing them with experimental data, while for $D$ and $D_s$ mesons, we compared our results with lattice data.  Moreover, in our model, the flavor symmetry is explicitly broken due to the different values of the quark masses and condensates.

Furthermore, for three-point functions, we studied the form factors $f_{+}(q^{2})$ of the following semileptonic decay processes, $D^{+} \to (\pi, K, \eta) l^{+} \nu_{l}$ and $D_{s}^{+} \to ( K, \eta) l^{+} \nu_{l}$ which associate with the exchange of a vector meson, and $V(q^{2})$ and $A_{1}(q^{2})$ of the $D_{(s)}^{+} \to K^{*} l^{+} \nu_{l}$ decays associated with the vector and axial vector meson exchange, respectively. The result of the form factor for $D^{+} \to \pi l^{+} \nu_{l}$,  $f_{+}(q^{2})$ shows excellent agreement with the experimental data, and it is comparable with lattice QCD and other theoretical approaches. Likewise, the normalized form factor $f_{+}(q^{2})$ of the $D_{(s)}$-to-kaon is very well consistent with the experimental and lattice data and has a better reproduction compared to other theoretical approaches; however, there is a $20\%$ discrepancy for $D_{(s)}\to K$ at zero momentum compare to experimental data.
Another semileptonic decay process is $D_{(s)}^{+} \to \eta l^{+} \nu_{l}$, similar to the form factors of the pion and kaon, the normalized $f_{+}(q^{2})$ for the $D^{+} \to \eta $ is compatible with data; however, a little deviation from the experimental data can be seen for $D_{s}^{+} \to \eta $. Finally, we predicted the vector form factors $V(q^{2})$ and $A_{1}(q^{2})$ for the decays $D \to K^{*}$ and $D_{s} \to K^{*}$. Our results agreed well with other approaches and lattice data at maximum-recoil $f_{+}(0)$ but increase dramatically at high momentum transfers, particularly for $D_{s} \to K^{*}$. These results gave us a signal that, there might be a missing dynamics at high momentum transfers, and in the future, we should deeply investigate these decay channels. 

In the future it would be interesting to extend the calculation of the semileptonic form factors of the B mesons, which contain the bottom quark using the holographic QCD model. Finally, we think that the model can be further improved by using an explicit expression of the dilaton profile, which respect the linear confinement and spontaneous symmetry breaking. We hope to dig down to these topics in the future work.

\section*{Acknowledgments}

This work is supported in part by the National Natural Science Foundation of China (NSFC) Grant Nos: 12235016, 12221005, 12147150, 12305136 and the Strategic Priority Research Program of Chinese Academy of Sciences under Grant No XDB34030000, the start-up funding from University of Chinese Academy of Sciences(UCAS), the start-up funding of Hangzhou Normal University under Grant No. 4245C50223204075, and the Fundamental Research Funds for the Central Universities. H. A. A. acknowledges the "Alliance of International Science Organization (ANSO) Scholarship For Young Talents" for providing financial support for the Ph.D. study.

 \addcontentsline{toc}{section}{References}
\end{document}